\documentclass[english,british,aps,reprint, pre]{revtex4-1}
\usepackage[T1]{fontenc}
\usepackage[latin9]{inputenc}
\usepackage{graphicx}
\usepackage{amsmath}

\makeatletter

\newenvironment{lyxlist}[1]
{\begin{list}{}
{\settowidth{\labelwidth}{#1}
 \setlength{\leftmargin}{\labelwidth}
 \addtolength{\leftmargin}{\labelsep}
 }}
{\end{list}}

\usepackage{braket}

\makeatother

\usepackage{babel}
\begin{document}

\title{A classical reactive potential for molecular clusters of sulphuric
acid and water}

\author{Jake L. Stinson$^{a}$, Shawn M. Kathmann$^{b}$ and Ian J. Ford$^{a}$$^{\ast}$
}

\address{$^{a}$Department of Physics and Astronomy and London Centre for
Nanotechnology, University College London, Gower Street, London, WC1E
6BT, United Kingdom}

\address{$^{b}$Physical Sciences Division, Pacific Northwest National Laboratory,
Richland, Washington 99352, United States}

\begin{abstract}
We present a two-state empirical valence bond (EVB) potential describing
interactions between sulphuric acid and water molecules and designed
to model proton transfer between them within a classical dynamical
framework. The potential has been developed in order to study the
properties of molecular clusters of these species, which are thought
to be relevant to atmospheric aerosol nucleation. The particle swarm
optimisation method has been used to fit the parameters of the EVB
model to density functional theory (DFT) calculations. Features of
the parametrised model and DFT data are compared and found to be in
satisfactory agreement. In particular, it is found that a single sulphuric
acid molecule will donate a proton when clustered with four water
molecules at $\mathrm{300}$ K and that this threshold is temperature
dependent.

\end{abstract}

\maketitle

\section{Introduction\label{sec:1Introduction}}

It has long been suspected that sulphuric acid plays an important
role in atmospheric particle nucleation as a consequence of its affinity
to water and its low volatility \citep{Zhang2012}, though ammonia
and organic species, as well as ions, are also likely to participate
\citep{Kulmala14}. Considerable experimental advances in characterising
nucleation phenomena in atmospherically relevant conditions, with
particular emphasis on sulphuric acid, have been reported in recent
years \citep{Kirkby2011,Riccobono14}. The interpretation of such
experimental data in order to understand behaviour over wider conditions
is critically dependent on calculations of the free energies of clusters
of the nucleating species. Such thermodynamic information is employed
within a well-established theoretical framework of cluster growth
and decay to predict rates of formation of stable clusters \citep{Ford2004,Kalikmanov-book2013}.
Although further processes such as cluster coalescence or removal
also play a role in particle formation \citep{McGrath12}, this kinetic
and thermodynamic framework lies at the heart of our understanding
of the phenomenon.

Cluster free energies, however, are not straightforward to calculate.
Simple models of cluster thermodynamic properties have repeatedly
been sought, but often such approaches rely on an extrapolation of
the properties of larger droplets down to clusters consisting of only
a few molecules \citep{girshick90,laaksonen94,kalikmanov95}. The
prime example of such a model is classical nucleation theory (CNT)
\citep{Vehkamaki,Ford2004}, based on the capillarity approximation.
Caution is necessary when using such models if they suggest that the
nucleation rate is sensitive to the properties of very small clusters,
since accuracy is not to be expected \citep{Brus2011}, though surprisingly
it is often observed.

More advanced models of molecular clusters can be used, of course,
though at greater cost in computational effort. At the most fundamental
level, molecular interaction models based on \emph{ab initio} quantum
mechanics are available. Indeed, very detailed studies have been conducted,
even taking account of the quantum nature of the lighter nuclei present
\citep{Kakizaki2009,Sugawara2011} in addition to that of the electrons.
In a structural study of the system that we consider in this paper,
for example, Kakizaki \emph{et al.} \citep{Kakizaki2009} concluded
that nuclear zero point motion in clusters of sulphuric acid and water
at 250 K gives rise to noticeably increased fluctuations and liquid-like
behaviour. Sugawara \emph{et al.} \citep{Sugawara2011} studied the
degree of hydration required for the dissociation of the sulphuric
acid molecule, at the same level of theory \citep{Arrouvel2005}.
Further evidence of small but significant effects of zero point motion
in similar clusters was provided by Stinson \emph{et al.} \citep{stinson14}.

However, when computing free energies using \emph{ab initio} methods,
the harmonic oscillator approximation is commonly employed, based
on identifying the lowest energy cluster configuration at zero temperature
and estimating the vibrational entropic contributions to the free
energy from a characterisation of the low temperature normal mode
spectrum. Such an approach is likely to be accurate at temperatures
where the cluster behaves as a vibrating solid-like structural network,
but would seem to be less appropriate for liquid-like systems. A more
general approach is then typically required, such as thermodynamic
integration \citep{Frenkel2001} where the free energy of a system
is compared with that of a better understood reference system at the
same temperature, through performing a series of canonical averages
with interpolating Hamiltonians, often using Monte Carlo (MC) methods.
Approaches based on nonequilibrium molecular dynamics (MD) simulation
have also received attention \citep{Ford2004c,Tang15}. Methods of
free energy computation are indeed rather numerous \citep{Pohorille07}.
In many systems of interest, however, \emph{ab initio} energy calculations
are too expensive, and modelling necessarily proceeds on a more coarse
grained, classical level.

A number of classical schemes have been employed to study clusters
of sulphuric acid and water molecules. In an early study Kusaka \emph{et
al.} \citep{Kusaka1998} developed a grand canonical MC model based
on rigid molecules and concluded that the clusters are highly non-spherical
and that bulk-like behaviour only emerges when there are at least
$240$ water molecules and $\mathrm{1-3}$ molecules of sulphuric
acid, or its dissociation product bisulphate, in the cluster. Later,
Kathmann and Hale \citep{Kathmann2001} presented a model based
on rigid water and sulphate molecules and a free $\mathrm{H}{}^{\delta+}$
ion, treated within an MC approach. Ding \emph{et al.} \citep{DingC.-G.}
developed a flexible bonding model for sulphuric acid, bisulphate,
hydronium and water species, which was employed by Toivola \emph{et
al.} \citep{Toivola2009b} to study a quasiplanar liquid-vapour interface
using clusters of $\mathrm{2000}$ molecules. Amongst other matters,
they found that when the sulphuric acid mole fraction is less than
0.1, the acid molecules lie at the cluster surface and that the cluster
structure is strongly dependent on the number of bisulphate ions present.

However, although mixtures of dissociated species were studied by
Toivola \emph{et al.} \citep{Toivola2009b}, the dynamics of the transfer
of protons between species was not considered in the model. It is
important to note that a classical potential suitable for free energy
computations for clusters of sulphuric acid and water ought to be
designed to describe the transfer of a proton from the acid to a water
molecule. Without such a capability, it is not possible to allow proton
transfer to take place naturally; instead, the degree of dissociation
would have to be fixed by hand in a given simulation.

This brings us to the aim of this study, which is to develop and parametrise
a classical potential that does offer this capability. Such a potential
would allow us to test the harmonic approximation in calculations
of the free energy in sulphuric acid/water clusters at relevant atmospheric
temperatures. Using a classical potential that can describe the dynamics
beyond structural vibrations is a crucial requirement.

There are a few classical schemes available which allow reactions
to occur. These include the Gaussian Approximation Potential (GAP)
\citep{Bartok2010} where a potential energy surface is constructed
by fitting a Gaussian basis set to reference data following a Bayesian
statistics procedure; the ReaxFF approach \citep{VanDuin2001} which
uses the bond order methodology; and the empirical valence bond (EVB)
model \citep{Kamerlin2011a} based on a superposition of underlying
classical states of the system.

The EVB methodology is attractive for use in the sulphuric acid/water
system since, as we shall see, it is based on classical potentials
for each species plus mixing terms, and is relatively straightforward
to implement into existing MC or MD schemes. The methodology was introduced
by Aqvist and Warshel \citep{Aqvist1993} and is reviewed by Kamerlin
and Warshel \citep{Kamerlin2011a}. It was first developed in order
to model proton transfer between hydronium and water species. Schmitt
and Voth \citep{Schmitt1998} designed a so-called multi-state
EVB (MS-EVB) model for the simulation of systems of water molecules
containing excess protons. This work was further developed into the
MS-EVB2 \citep{Day2002} and the MS-EVB3 \citep{Wu2008} models. Our
strategy is to use the MS-EVB3 model as a framework for constructing
an EVB model for the sulphuric acid and water system. As a simplification,
we limit the multiplicity of states to two (i.e. the proton may be
attached either to a bisulphate or to the nearest appropriate water
molecule) making it a two level EVB approach. Multiple potential proton
transfers within the system are accommodated by employing the self-consistent
iterative MS-EVB (SCI-MS-EVB) model, developed by Wang and Voth
\citep{Wang2005} for water with excess protons. A recent summary of the EVB model in the context of modelling proton transfer in a water network is given by Knight and Voth  \citep{Knight12}.

We describe the basis of an EVB model in Section \ref{sec:2EVB-Method}.
Section \ref{sec:3EVB-model-for-SA-WA} is an account of the specific
elements of the model developed here for sulphuric acid and water
mixtures. A demonstration of the model is given in Section \ref{sec:5Results}
and in Section \ref{sec:6Conclusion} we summarise and discuss future
applications. The Appendix describes some technical aspects of the
particle swarm optimisation (PSO) fitting procedure used to parametrise
the model against \emph{ab initio} calculations at the DFT level of
theory.

\section{The EVB method\label{sec:2EVB-Method}}

\subsection{EVB overview}

\begin{figure*}
\begin{centering}
\includegraphics[width=1\textwidth]{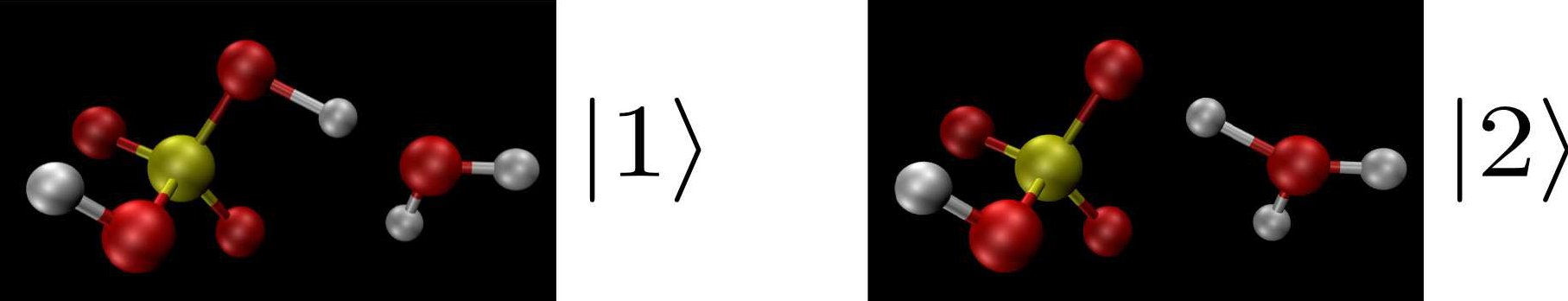}
\par\end{centering}

\protect\caption{Illustration of two possible basis states of bonding for the same
atomic configuration. State $\Ket{1}$ is composed of a sulphuric
acid and a water molecule whereas state $\Ket{2}$ contains bisulphate
and hydronium ions. The EVB approach provides an interpolation between
the energy surfaces appropriate to each case. \label{fig:1BasisSetExample}}
\end{figure*}

In an EVB model the potential energy of the reactive
system is constructed as an interpolation between energy surfaces
defined for specific choices of bonding pattern. This is done in a
fashion reminiscent of the quantum mechanics of a multilevel system,
though it should be emphasised that the dynamics considered are entirely
classical. The system is represented mathematically as a superposition
of basis states, each representing a possible bonding pattern. For
example, Figure \ref{fig:1BasisSetExample} illustrates two basis
states of a system corresponding to the possibility in this case that
the central hydrogen atom can form part of either a sulphuric acid
molecule on the left or a hydronium ion on the right.

Once the basis states of the system are specified, a matrix $\mathcal{H}$
that resembles a Hamiltonian is constructed as follows:

\begin{equation}
\mathcal{H}_{ij}=\Bra{i}H\Ket{j},\label{eq:HConstruction}
\end{equation}

\noindent where $\Bra{i}H\Ket{i}$ is the potential energy of the
system according to the bonding pattern defined by $\Ket{i}$ and
an associated classical potential. $\Bra{i}H\Ket{j}$, where $i\neq j$,
represents the coupling between basis states $\Ket{i}$ and $\Ket{j}$
responsible for mixing. In practice, $\Bra{i}H\Ket{j}$ is specified
by an empirical function chosen to reproduce the behaviour of a higher
level theoretical model.

Once the $\mathcal{H}$ matrix has been constructed, the eigenvectors
and eigenvalues are found. The eigenvector $\Ket{\Psi}=\sum_{i}c_{i}\Ket{i}$
with the lowest eigenvalue represents the ground state of the system,
with coefficients $c_{i}$ specifying the appropriate superposition
of the basis states. The energy is given by $E=\boldsymbol{c}^{T}\mathcal{H}\boldsymbol{c}$
and the forces can be computed via the Hellmann-Feynman theorem

\begin{equation}
\mathbf{F}_{n}=-\left\langle {\Psi}\right|\frac{\partial H}{\partial\boldsymbol{x}_{n}}\left|{\Psi}\right\rangle =-\sum_{i,j}c_{i}c_{j}\frac{\partial\Bra{i}H\Ket{j}}{\partial\boldsymbol{x}_{n}},\label{eq:HellmannFeynmanTheorem}
\end{equation}

\noindent where $\boldsymbol{x}_{n}$ and $\mathbf{F}_{n}$ indicate
the position and the force, respectively, for atom $n$.

\subsection{The MS-EVB3 model of water systems with an excess proton}

The MS-EVB3 model \citep{Wu2008} was originally designed for simulating
systems of water molecules with one excess proton. The key process
of interest is the mobility of the excess proton across hydrogen bonds
formed between water and hydronium. Diagonal elements of the matrix
$\mathcal{H}$ are specified by a suitable potential for water and
hydronium bonding patterns. Off-diagonal components of $\mathcal{H}$
are calculated in the following way:
\begin{equation}
\Bra{i}H\Ket{j}_{i\neq j}=(V_{{\rm const}}^{ij}+V_{{\rm ex}}^{ij})\cdot A(R_{{\rm OO}},\mathbf{q}),
\end{equation}
where $R_{{\rm OO}}$ is the oxygen-oxygen distance in the hydrogen
bond that includes the transferring proton, $V_{{\rm const}}^{ij}$
is a constant and $V_{{\rm ex}}^{ij}$ is a representation of Coulombic
interactions between the ${\rm H}_{5}{\rm O}_{2}^{+}$ Zundel cation
consisting of the hydronium and water between which the proton is
considered to be shared, and the remaining waters. It is given by

\begin{equation}
V_{{\rm ex}}^{ij}=\sum_{m}^{7}\sum_{k}^{N_{{\rm H_{2}O}}-1}\sum_{n_{k}}^{3}\frac{q_{n_{k}}^{{\rm H_{2}O}}q_{m}^{ex}}{R_{mn_{k}}},\label{eq:VijExMSEVB3}
\end{equation}
where $m$ is a label for the seven atoms in the Zundel cation, $k$
is a label for the remaining water molecules ($N_{{\rm H_{2}O}}$
is the total number of water molecules in the system) and $n_{k}$
labels the three constituent atoms of each water molecule. $q_{n_{k}}^{{\rm H_{2}O}}$
are the partial charges for the constituent atoms in the water molecule
and $q_{m}^{ex}$ are partial charges describing the Zundel cation.
$R_{mn_{k}}$ is the distance between atoms labelled by $m$ and $n_{k}$.

\begin{table}
\caption{Parameters specifying the off-diagonal matrix elements of the Hamiltonian used in the MS-EVB3 model of water-hydronium transfer \citep{Wu2008}; $\alpha$, $\beta$ and $\gamma$ from \cite{Wu2008}, all others from \cite{Schmitt1998}. \label{tab:MSEVB3Values}}
{\begin{tabular}{ccccc}
\toprule
$V_{{\rm const}}^{ij}$ & -23.1871874  $\mathrm{kcal/mol}$ &  & $\beta$ & 4.5  $\mathrm{\textrm{\AA}^{-1}}$\tabularnewline

$\gamma$ & 1.85  $\mathrm{\textrm{\AA}^{-2}}$ &  & $R_{{\rm OO}}^{0}$ & 3.1  $\mathrm{\textrm{\AA}}$\tabularnewline  $P$ & 0.2327246   &  & $P'$ & 10.8831327  \tabularnewline  ${\rm k}$ & 9.562153  $\mathrm{\textrm{\AA}^{-2}}$ &  & $\alpha$ & 15  $\mathrm{\textrm{\AA}}^{-1}$\tabularnewline  $D_{{\rm OO}}$ & 2.94  $\mathrm{\textrm{\AA}}$ &  & $r_{{\rm OO}}^{0}$ & 1.8136426  $\mathrm{\textrm{\AA}}$\tabularnewline
\botrule
\end{tabular}}
\end{table}

The function $A(R_{{\rm OO}},\mathbf{q})$ has the form

\begin{eqnarray}
\begin{array}{c}
A(R_{{\rm OO}},\mathbf{q})=\exp(-\gamma\mathbf{q}^{2})\cdot\left\{ 1+P\exp\left[-{\rm k}(R_{{\rm OO}}-D_{{\rm OO}})^{2}\right]\right\} \\
\\
\times\left\{ \frac{1}{2}\left\{ 1-\tanh\left[\beta(R_{{\rm OO}}-R_{{\rm OO}}^{0})\right]\right\} +\right.\\
\\
\left.P'\exp\left[-\alpha(R_{{\rm OO}}-r_{{\rm OO}}^{0})\right]\right\} ,\\
\\
\end{array}\label{eq:MSEVBOFFD}
\end{eqnarray}

\noindent where $\gamma$, $P$, ${\rm k}$, $D_{{\rm OO}}$, $\beta$,
$R_{{\rm OO}}^{0}$, $P'$, $\alpha$ and $r_{{\rm OO}}^{0}$ are
empirical parameters (see Table \ref{tab:MSEVB3Values}) and $\mathbf{q}=\frac{1}{2}(\mathbf{r}_{{\rm O}1}+\mathbf{r}_{{\rm O}2})-\mathbf{r}_{{\rm H}}$
where $\mathbf{r}_{{\rm O}1}$, $\mathbf{r}_{{\rm O}2}$ and $\mathbf{r}_{{\rm H}}$
are the positions of the two oxygens and hydrogen, respectively, that
participate in the hydrogen bond. The model is somewhat empirical
in construction but was designed to have a number of desirable features
\citep{Wu2008}.

\subsection{The SCI-MS-EVB procedure\label{sub:Multi-proton-extension-paper}}

The EVB method has the capacity to model systems of $n>1$ hydroniums
(excess protons) with water, but this extension increases the size
of $\mathcal{H}$ to order $m^{n}$ where $m$ is the size of the
basis set for a single proton system. The SCI-MS-EVB model was developed
by Wang and Voth \citep{Wang2005} to improve the scaling of
the procedure with respect to $n$. This is achieved by treating the
system as $n$ single excess protons in the presence of an environment
with a given state of bond mixing. Employing a single proton EVB methodology,
the state of mixing of the entire system is then given by a set of
$n$ eigenvectors each of length $m$. The eigenvectors are corrected
iteratively as the effects of the other excess protons on the Hamiltonian
matrix for each of the Zundel cations are taken into account. Essentially,
the environment affects how the competing energy surfaces are specified.
The problem of analysing one matrix of order $m^{n}$ is hence reduced
to the consideration of $n$ matrices of size $m$. The number of
iterations required to reach convergence in energy and forces is typically
small. We similarly use this procedure for our system.

\section{EVB model for sulphuric acid and water\label{sec:3EVB-model-for-SA-WA}}

In the following subsections we describe in some detail the construction
of an EVB model for a system containing sulphuric acid, bisulphate,
hydronium and water species. It is a complex narrative, but a summary
of the procedure is provided at the end in Subsection \ref{sub:Model-Overview}.
In the discussion a naming convention is used where the \textbf{\tt{ground state}}
refers to the bonding pattern that has been identified as geometrically
the most `natural', based upon the configuration of the atoms. \textbf{\tt{Excited states}}
are bonding configurations that are identical to the \textbf{\tt{ground state}}
except for the repositioning of one bond as a consequence of a proton
transfer. Therefore, it is possible to identify an \textbf{\tt{excited state}
}based upon the\textbf{ \tt{ground state} }and the two species involved
in a proton transfer. The species are then referred to as a \textbf{\tt{donor}}
(the molecule to which the hydrogen atom is bonded in the \textbf{\tt{ground state}}
and which can either be a sulphuric acid molecule or a hydronium ion),
and an \textbf{\tt{acceptor}} (the molecule to which the hydrogen
atom is bonded in the \textbf{\tt{excited state}} and which can either
be a bisulphate or a water). For clarity, the terms neutral and ionised
are used to identify the state of the sulphur-bearing species (either
as a neutral sulphuric acid or an ionised bisulphate).

\subsection{Basis set size\label{sub:Basis-set-size}}

The basis set size $m$ for each transferable proton was chosen to
be two, in order to keep the model as simple as possible, in line
with the SCI-MS-EVB approach implemented for water and described in
Section \ref{sub:Multi-proton-extension-paper}. This approach allows
each donor and acceptor species in the system to be involved in a
maximum of one proton transfer in a given configuration, but it is
assumed that the possibility that such a species might be involved
in two proton transfers simultaneously can be safely neglected. The
small basis set leads to occasional ambiguities concerning which proton
is considered to be shared but a modification of the off-diagonal term
specified in Eq. (\ref{eq:MSEVBOFFD}) has been made to limit the
impact of these issues and is described in Section \ref{subsub:qdependence}.

 Our focus here has been to develop the simplest possible scheme of the process of proton transfer between sulphuric acid and water, and for ease of implementation we have used a similar two-state model for the water network. Larger basis sets have previously been employed in studies of the water/hydronium system in order to capture effects such as the formation of highly correlated structures such as the H$_9$O$_4$$^+$ complex discussed in \citep{Knight12}. Our approach is less refined at present, but clearly a more elaborate model of state mixing within the water network could be implemented at a later stage alongside the two-state model of acid/water mixing in order to capture the finer details of proton transfer in the water sector. Our principal focus has been to describe the reactive behaviour of sulphuric acid in clusters containing relatively few water molecules.

\begin{figure*}
\begin{centering}
\includegraphics[width=1\textwidth]{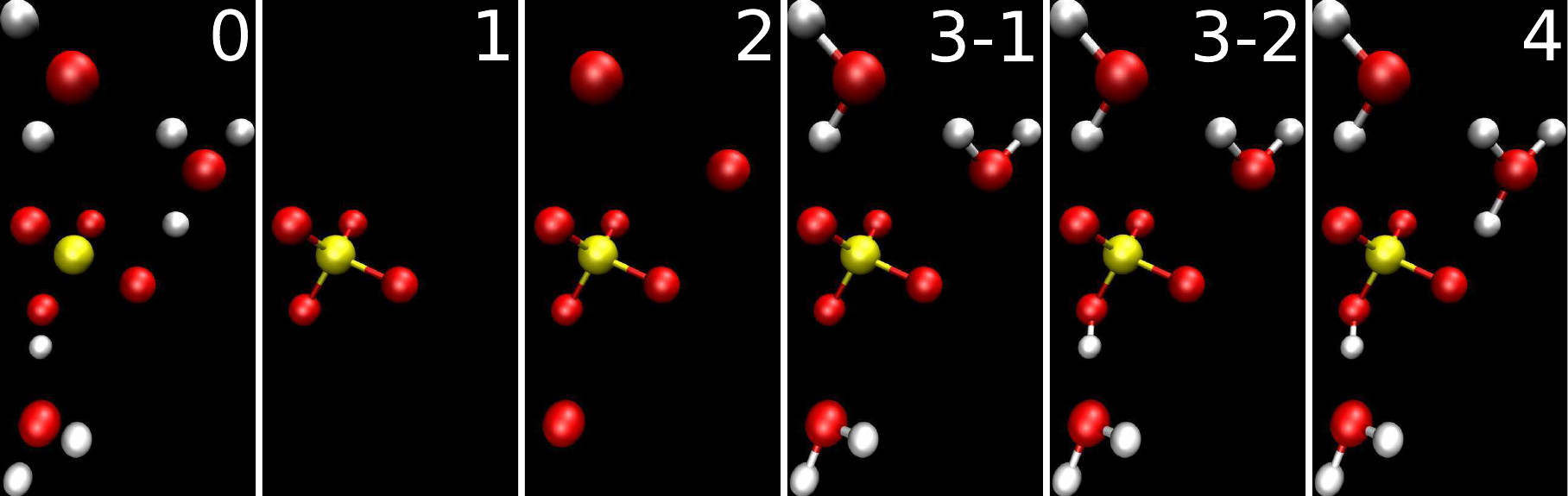}
\par\end{centering}

\protect\caption{\label{fig:State-selector-Image}The ground state selector algorithm.
Image zero shows the positions of atoms, and in subsequent steps bonds
are inserted according to the procedure described in Section \ref{sub:State-selector-algorithm}.
Step three is shown in two parts; first, the attachment of two hydrogen
atoms to each oxygen and second, the attachment of one hydrogen
atom to each sulphate.}
\end{figure*}

\subsection{Algorithms}

A central part of the EVB methodology is the construction of appropriate
basis states of the system, which is nontrivial as reactions can cause
atoms to associate with different molecules in the course of a simulation.
This section describes two algorithms which are applied at every time
step in order to identify the ground state and excited states of the
system by geometric arguments.

\subsubsection{Ground state selector algorithm\label{sub:State-selector-algorithm}}

An algorithm has been designed which constructs the \textbf{\tt{ground state}}
from a list of atomic positions. For convenience, four molecular lists
are defined and are denoted by \textbf{\tt{SA}} (sulphuric acid),
\textbf{\tt{BS}} (bisulphate), \textbf{\tt{HY}} (hydronium) and
\textbf{\tt{WA}} (water) and an assignment refers to the identification
of an atom as a constituent of a particular molecule. The algorithm
is performed in the following way:
\begin{enumerate}
\item The four oxygens closest to each sulphur atom are identified. They
are assigned to a sulphuric acid molecule (along with the sulphur
atom) and placed in the \textbf{\tt{SA}} list.
\item The remaining oxygen atoms are assigned to hydronium ions and added
to the \textbf{\tt{HY}} list.
\item Minimum numbers of hydrogen atoms are assigned to each molecular species.
For each member of the \textbf{\tt{HY}} list the \emph{two} hydrogen
atoms closest to the associated oxygen are identified and assigned
to the list. The same process is used to assign \emph{one} hydrogen
to each member of the \textbf{\tt{SA}} list.
\item The remaining hydrogen atoms are placed (in no particular order) in
a temporary list named \textbf{\tt{H}}. The closest oxygen to each
of these hydrogens is identified and the hydrogen is assigned to the
molecular species of which the oxygen is a constituent. The oxygen
cannot be part of a molecule which has already accepted one of these
remaining hydrogens and cannot be a member of the \textbf{\tt{SA}}
list that was assigned a bond to a hydrogen atom in step 3. If there
is an attempt to assign a hydrogen where the bond length is greater
than $1.2\textrm{\AA}$ then this assignment is rejected and the atom
is moved to the top of the \textbf{\tt{H}} list. If the \textbf{\tt{H}}
list is rearranged at any time, then step 4 is immediately restarted
using the modified \textbf{\tt{H}} list. The hydrogens which went
through a rejected assignment are no longer subject to the $1.2\textrm{\AA}$
constraint, but if the bond length is over $2\textrm{\AA}$ the hydrogen
is again returned to the top of the \textbf{\tt{H}} list. If problems
with assignment persist, then the algorithm is run with the bond length
check turned off. In practice, there is only need for these checks
in the circumstances of proton transfer events, in which case the
two state EVB mixing between the states will ensure that the appropriate
forces are applied regardless of the assignment of the molecular species.
This step is completed when all hydrogens in the \textbf{\tt{H}}
list have been assigned. The procedure followed in this step was found
to work smoothly in trials of the model.
\item Any sulphuric acid molecule in the \textbf{\tt{SA}} list for which
only one hydrogen atom has been assigned is moved to the \textbf{\tt{BS}}
list. Similarly, any hydronium molecule in the \textbf{\tt{HY}} list
that does not have three assigned hydrogen atoms is moved to the \textbf{\tt{WA}}
list.
\end{enumerate}
Figure \ref{fig:State-selector-Image} illustrates the steps of the
algorithm as they are performed on a system. At the end of the procedure,
the \textbf{\tt{SA}}, \textbf{\tt{BS}}, \textbf{\tt{HY}} and \textbf{\tt{WA}}
lists represent the \textbf{\tt{ground state}} of the system.

\begin{figure}
\begin{centering}
\includegraphics[width=1\columnwidth]{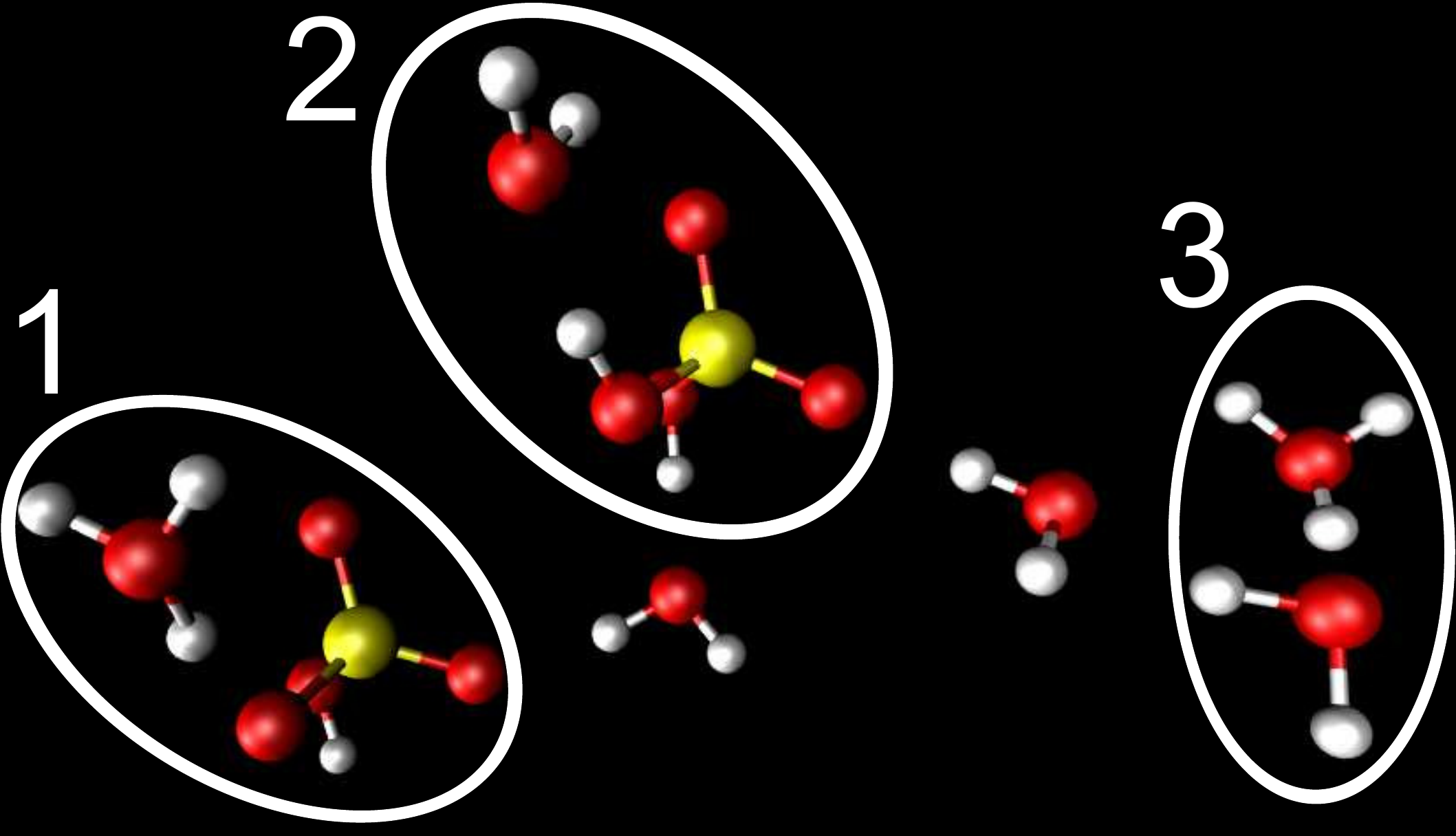}
\par\end{centering}

\protect\caption{\label{fig:ExcitedStateImage}A simultaneous consideration of three
potential proton transfer events. The groups labelled 1, 2 and 3 refer
to {[}hydronium/bisulphate{]}, {[}sulphuric acid/water{]} and {[}hydronium/water{]}
proton transfer, respectively. One \textbf{\tt{ground state}} (the
bonding pattern shown) and three \textbf{\tt{excited states}} (where
the central hydrogen in each group is bonded differently) have therefore
been identified.}
\end{figure}

\subsubsection{Excited state identification algorithm\label{sub:Excited-basis-states}}

As stated earlier, an \textbf{\tt{excited state}} is defined as a
change in the bonding assignment of one hydrogen in the \textbf{\tt{ground state}}
of a given configuration. An algorithm identifies the\textbf{ \tt{excited states}}
in a system using the following steps:
\begin{enumerate}
\item The shortest distance between a hydrogen belonging to a \textbf{\tt{donor}}
species, and an oxygen belonging to an \textbf{\tt{acceptor}} species
is identified. The oxygen in a bisulphate ion which has an attached
hydrogen is not considered here. If this distance is less than $2\textrm{\AA}$
then an\textbf{ \tt{excited state}} may be constructed where the
hydrogen is reassigned to the \textbf{\tt{acceptor}} species.
\item Step 1 is repeated until either the shortest distance identified is
greater than $2\textrm{\AA}$ in length or there are no further \textbf{\tt{acceptor}
}or\textbf{ \tt{donor}} molecules for which it is possible to construct
an $\mathrm{O-H}$ separation distance.
\end{enumerate}
The ground state selector and the excited state identification algorithms
produce one \textbf{\tt{ground state}} and $n$ \textbf{\tt{excited states}}
where $n$ is the number of potential proton transfer events in the
configuration under consideration. When $n=0$ the system is in a
non-reactive configuration. In the configuration shown in Figure \ref{fig:ExcitedStateImage},
there are three groups where the bonding of a proton is unclear (ringed),
such that a \textbf{\tt{ground state}} and three \textbf{\tt{excited states}}
have been identified in this system.

\subsection{Diagonal terms\label{sub:Underlying-classical-potential}}

The potentials needed to specify the diagonal elements of $\mathcal{H}$
are those developed by Loukonen \emph{et al.} \citep{Loukonen2010},
and the SPC/EF potential is used for the water molecules \citep{Lopez-Lemus08}.
The hydronium energy is represented by a set of harmonic angle potentials
and Morse bond potentials as follows:
\begin{equation}
V_{{\rm hyd}}=\frac{1}{2}\sum_{i=1}^{3}k_{\theta}\left(\theta_{i}-\theta_{0}\right)^{2}+\sum_{j=1}^{3}D\left[1-{\rm e}^{-\alpha(r_{j}-r_{0})}\right]^{2},\label{eq:Chap6_hydronium_potenial}
\end{equation}
where $i$ is summed over the three hydrogen-oxygen-hydrogen angles
and $j$ over the three oxygen-hydrogen bonds. The $k_{\theta}$ and
$\theta_{0}$ parameters are taken from Loukonen \emph{et al.} \citep{Loukonen2010}.
The Morse potential is similar in form to the model used in Wu \emph{et
al.} \citep{Wu2008}, and employs the same value of $D$, but the
$\alpha$ parameter has been changed so that the second order Taylor
expansion around $r_{0}$ matches the strength of the harmonic spring
used to represent the OH bond in Loukonen \emph{et al.} \citep{Loukonen2010},
namely $\alpha=2.327\textrm{\AA}^{-1}$. The use of a Morse potential
rather than a harmonic spring was found to improve the match between
our model of hydronium and that used in the original MS-EVB3 model\@.

One of the diagonal terms is augmented by an energy shift, $\Delta$,
to account for the difference in zero temperature ground state energy
between the sulphuric acid and water bonding pattern and the bisulphate
and hydronium version. Ding \emph{et al.} \citep{DingC.-G.} calculated
the value of this parameter to be $144.0\:\mathrm{kcal/mol}$ ($602.5\:\mathrm{kJ/mol}$),
but for our purposes it was considered to be a free parameter and
is fitted to higher level theory data.

\subsection{Off-diagonal terms\label{sub:Off-Diagonal}}

The form chosen for the off-diagonal terms in the Hamiltonian describing
sulphuric acid-water proton transfer is based upon the MS-EVB3 model
(Eq. (\ref{eq:MSEVBOFFD})), with two modifications. These relate
to the expression involving the $\mathbf{q}$ parameter (Section \ref{subsub:qdependence})
and the method for calculating $V_{{\rm ex}}^{ij}$ (Section \ref{subsub:Intermediate-charges-1}).

\subsubsection{The $\mathbf{q}$ dependence\label{subsub:qdependence}}


 There is a problem with using a limited basis set, especially when there are long-range off-diagonal terms in the Hamiltonian matrix. If configurational evolution brings about a change in identity of the species that might accept the proton from a given \textbf{\tt{donor}} species, then the replacement of one energy surface by another will bring about an abrupt jump in energy that will need to be damped out by the thermostat. However, we have designed the two-state EVB scheme to avoid such jumps as much as possible. Mixing between bonding patterns will only come about when the proton in question occupies a specific region between the \textbf{\tt{donor}} and \textbf{\tt{acceptor}} species. Outside this region, the interactions revert to those of a single pattern of bonding, and crucially, the energy surface is continuous at the boundary of the region. This is encoded in the specification of off-diagonal terms which vanish for proton positions outside an ellipsoidal volume centred on the mid-point between the donating and accepting oxygen atoms.

Specifically, the factor in
Eq. (\ref{eq:MSEVBOFFD}) involving $\mathbf{q}$ becomes
\begin{equation}
\exp(-\gamma\mathbf{q}^{2})\rightarrow\begin{cases}
\left[\frac{\exp(-(\gamma\mathbf{q}_{0}^{2}-1))-1}{{\rm e}-1}\right] & \mathbf{q}_{0}^{2}\leq\gamma^{-1}\\
0 & \mathbf{q}_{0}^{2}>\gamma^{-1}
\end{cases}\label{eq:gammaMod}
\end{equation}
where $\mathbf{q}_{0}^{2}$ is related to $\mathbf{q}$ in the following
way. The $\mathbf{q}$ vector is expressed in a new orthogonal coordinate
system $\mathbf{q}^{\prime}$ in which the $x$-axis is parallel to
$\mathbf{r}_{{\rm OO}}$, where $\mathbf{r}_{{\rm OO}}=\mathbf{r}_{{\rm O}1}-\mathbf{r}_{{\rm O}2}$
is the vector separation between oxygens. The other two axes take
arbitrary orientations. $\mathbf{q}_{0}^{2}$ is then defined as $\mathbf{q}_{0}^{2}=\left(q'_{x}/\tau_{x}\right)^{2}+\left(q'_{y}/\tau_{yz}\right)^{2}+\left(q'_{z}/\tau_{yz}\right)^{2}$
where the $\tau_{x}$ and $\tau_{yz}$ are unitless scaling factors
representing semi-major axes of an ellipsoid with respect to the $x$
axis, and the $y$ and $z$ axes, respectively. In effect, a hydrogen
atom experiences mixed bonding within an ellipsoidal volume between
the two appropriate oxygen atoms. We find that this eliminates occasional ambiguities with regard to
which oxygen pair is appropriate for the mixed bonding experienced
by a particular hydrogen, introducing discontinuities in
the potential energy landscape. The two variables $\tau_{x}$ and
$\tau_{yz}$ are treated as additional fitting parameters. The form
of Eq. (\ref{eq:gammaMod}) ensures two features: the expression is
equal to unity at $\mathbf{q}_{0}^{2}=0$ and goes to zero on the
surface of the ellipsoid at $\mathbf{q}_{0}^{2}=\gamma^{-1}$, beyond
which the off-diagonal term vanishes to enforce ordinary unmixed
behaviour outside the ellipsoid.

 In principle, a  jump in energy can still occur if ellipsoidal mixing regions between two donor-acceptor pairs come into overlap as a consequence of configurational change. Such situations can occasionally arise during the molecular dynamics, but we have restricted the regions of mixing, through the parameters $\tau_x$ and $\tau_{yz}$, specifically to avoid this. For most of the time there is no ambiguity in the choice of basis set to use for a configuration and few occasions when a jump in energy occurs according to our scheme. These jumps are unphysical, and a breakage of NVE conditions, but are the price to pay for the simplicity of the implementation.

\subsubsection{Interpolated charges\label{subsub:Intermediate-charges-1}}

The $V_{{\rm ex}}^{ij}$ parameter in Eq. (\ref{eq:VijExMSEVB3})
is also modified since the charges provided by the MS-EVB3 model for
the calculation of this term are specific to the Zundel cation and
do not transfer to a sulphuric acid and water model. We recall that
it represents external electrostatic interactions for the group of
atoms involved in the mixing of bonding patterns. Our approach is
to interpolate the charges of these atoms according to the $c_{i}$
coefficients of the mixture. The procedure is as follows:
\begin{enumerate}
\item Initially the charges are assumed to be $q=\frac{1}{2}(q^{gs}+q^{es})$,
where $q^{gs}$ and\textbf{ }$q^{es}$ are the partial charges for
that atom according to the classical potential provided, in accordance
with whether the bonding is identified as the \textbf{\tt{ground state}
}or the \textbf{\tt{excited state}}, respectively.
\item A self-consistent iteration is performed where \foreignlanguage{english}{$q$}
is updated according to $q=c_{gs}^{2}q^{gs}+c_{es}^{2}q^{es}$ where
$c_{gs}$ and $c_{es}$ are the eigenvector coefficients for the \textbf{\tt{ground state}
}and \textbf{\tt{excited state}}, respectively, obtained from the
two-level Hamiltonian describing the mixing.
\end{enumerate}
This procedure is performed alongside a separate self-consistent iterative
process described in the next section, the purpose of which is to
allow consideration of multiple proton transfers.

\subsection{Multiple proton procedure \label{sub:Multiple-proton-procedure}}

The modelling of systems undergoing multiple proton transfers follows
the SCI-MS-EVB procedure developed by Wang and Voth \citep{Wang2005}.
Figure \ref{fig:ExcitedStateImage} shows a system where this extension
is required as it has three \textbf{\tt{excited states}} available.
The starting point in this procedure is to neglect overlap between
two different excited states (i.e. $\bra{i}H\ket{j}=0$ where $i$
and $j$ represent two different excited states). This allows for
the reduction of the EVB Hamiltonian matrix of the system, which is
of order $n_{d}^{n_{c}}$, to $n_{d}$ matrices of order $n_{c}$
where $n_{d}$ is the number of donor molecules being considered and
$n_{c}$ is the number basis states considered per donor molecule.
An iterative procedure is then used to correct these matrices for
the $\bra{i}H\ket{j}$ contributions. The method can be described
as follows:
\begin{enumerate}
\item We determine the eigenvector for each $2\times2$ matrix $\mathcal{H}$
corresponding to the \textbf{\tt{ground state} }and an\textbf{ \tt{excited state}}.
For clarity, $n$ is  used as a label such that $\mathbf{c}_{n}$
and $\mathcal{H}_{n}$ are defined for the $n^{{\rm th}}$\textbf{
\tt{excited state}}. In addition, $n$ is used to label the \textbf{\tt{donor}}
and \textbf{\tt{acceptor} }species that define the \textbf{\tt{excited state}}.
\item Each $\mathcal{H}_{n}$ is then corrected for the effect of \textbf{\tt{excited state}
$m$} where $m\neq n$. This is performed by updating the intermolecular
energy contributions to the diagonal terms in $\mathcal{H}_{n}$,
i.e. $E_{{\rm inter}}^{gs}\rightarrow\left(c_{gs}^{2}E_{{\rm inter}}^{gs}+c_{es}^{2}E_{{\rm inter}}^{es}\right)$
where $c_{gs}^{2}$ and $c_{es}^{2}$ are the squared coefficients
of the $\mathbf{c}_{m}$ eigenvector representing the associated \textbf{\tt{ground state}}
and \textbf{\tt{excited state}}. $E_{{\rm inter}}^{gs}$ and $E_{{\rm inter}}^{es}$
are the intermolecular energy contributions resulting from interactions
between the \textbf{\tt{acceptor}}/\textbf{\tt{donor}} pair $n$
and the \textbf{\tt{acceptor}}/\textbf{\tt{donor}} pair $m$ in
the \textbf{\tt{ground state}} and \textbf{\tt{excited state}} of
$m$ respectively. The expression $V_{{\rm ex}}^{ij}$ in the off-diagonal
term is updated in the same fashion; this includes recalculating the
charges \foreignlanguage{english}{$q$} according to the procedure
in Section \ref{subsub:Intermediate-charges-1}. The $\mathbf{c}_{n}$
values are then recalculated. This allows the coefficients\textbf{
$\mathbf{c}_{n}$ }for the\textbf{ \tt{acceptor}}/\textbf{\tt{donor}}
pair $n$\textbf{ }to be corrected for the intermolecular interactions
arising from \textbf{\tt{acceptor}}/\textbf{\tt{donor}} pair $m$.
This step is repeated until each $\mathbf{c}_{n}$ vector has been
corrected for each \textbf{\tt{acceptor}}/\textbf{\tt{donor}} pair
$m$ (where $m\ne n$).
\item Step $2$ is repeated until all $\mathbf{c}_{n}$ eigenvectors have
converged. This is tested by defining $C_{{\rm sum}}=\sum_{n}(\mathbf{c}_{n}^{{\rm old}}-\mathbf{c}_{n}^{{\rm new}})^{2}$
where $\mathbf{c}_{n}^{{\rm old}}$ and $\mathbf{c}_{n}^{{\rm new}}$
are the (normalised) $\mathrm{\mathbf{c}}_{n}$ eigenvectors calculated
before and after step $2$ is performed. The system is considered
to be converged when $C_{{\rm sum}}<10^{-5}$, or when step 2 has
been cycled 10 times. In practice only a few iterations are required.
\end{enumerate}

\subsection{Energy and force calculation\label{sub:Force-calculation}}

Once the self-consistency iterations have been performed, the energy
of the system can be computed in the same fashion as in the SCI-MS-EVB
method. Eq. (13) in reference \citep{Wang2005}  gives a deconstruction
of the total energy of a system containing two excess protons in the
form

\begin{equation}
E_{{\rm total}}=E_{AA}+E_{BB}+E_{AB}+E_{AR}+E_{BR}+E_{RR},\label{eq:RefTotalEnergy}
\end{equation}

\noindent where $E_{{\rm total}}$ refers to the total energy of a
system according to the EVB method. $A$ and $B$ refer to separate
\textbf{\tt{acceptor}}/\textbf{\tt{donor} }pairs (see Figure 1 in
reference \citep{Wang2005}). $E_{AA}$ and $E_{BB}$ refer to the
independent energy contributions of the \textbf{\tt{acceptor}}/\textbf{\tt{donor}
}pairs and includes their off-diagonal contributions. $E_{AB}$ describes
the energy contribution due to interactions between $A$ and $B$.
$R$ refers to the rest of the system, and there are three further
contributions due to interactions between $A$ and $R$, $B$ and
$R$ and the independent energy contribution of $R$, referred to
as $E_{AR}$, $E_{BR}$ and $E_{RR}$ respectively. Eq. (\ref{eq:RefTotalEnergy})
can be generalised to $E_{{\rm total}}=E_{RR}+\sum_{n}\left(E_{nn}+E_{nR}+\sum_{m\ne n}E_{nm}\right)$
where $n$ and $m$ refer to the identified \textbf{\tt{donor}}/\textbf{\tt{acceptor}
}pairs. Once $E_{{\rm total}}$ has been constructed it simply remains
to apply the Hellmann-Feynman theorem to determine the force acting
upon each atom within the system.

\subsection{Model overview\label{sub:Model-Overview}}

In summary, the model is an algorithm consisting of the following
series of steps:
\begin{enumerate}
\item Construct a\textbf{ \tt{ground state}} from a set of atomic positions
(Section \ref{sub:State-selector-algorithm}).
\item Identify the \textbf{\tt{excited states}} (Section \ref{sub:Excited-basis-states}).
\item Calculate the on-diagonal and off-diagonal terms of the $2\times2$
matrix $\mathcal{H}$ for each \textbf{\tt{excited state} }(Sections
\ref{sub:Underlying-classical-potential} and \ref{subsub:qdependence}).
\item Optimise the $\mathbf{c}_{i}$ vectors for each \textbf{\tt{excited state}}
by performing a self-consistent iterative procedure which revises
the off-diagonal terms and allows multiple proton transfers to be
accommodated (Sections \ref{subsub:Intermediate-charges-1} and \ref{sub:Multiple-proton-procedure}).
\item Calculate the energy of the system and the forces acting upon each
atom (Section \ref{sub:Force-calculation}).
\end{enumerate}
Once the parameters of the model have been selected, the procedure
can be implemented in a classical molecular dynamics code, and we
have done this using the DL\_POLY $\mathrm{4.03}$ package \citep{Todorov2006}.

\begin{figure}
\begin{centering}
\includegraphics[width=1\columnwidth]{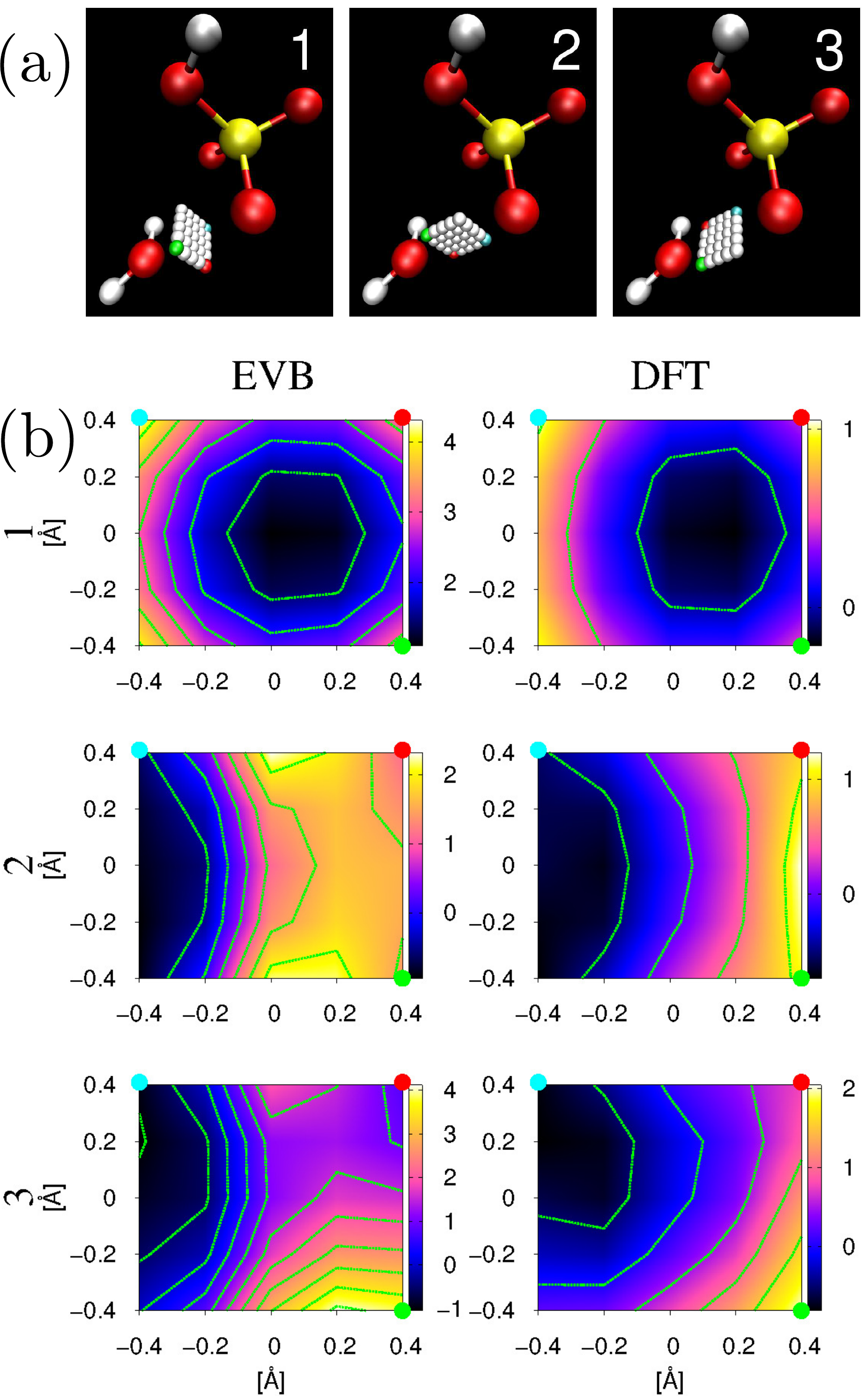}
\par\end{centering}

\begin{centering}
\protect\caption{Sets of positions of a hydrogen atom in spatial regions of mixed bonding
(a), together with corresponding contour plots of the potential energy
in units of eV (b), for the I-n configuration (labelled according
to reference \citep{Re1999}) of a sulphuric acid and a water molecule.
The centre of the contour plots lies at the mid-point between the
two participating oxygens and the labels 1, 2 and 3 in (b) refer to
the arrays of points shown in (a). The cyan, red and green spheres
in (a) match the equivalent top left, top right and the bottom right
of each plot in (b), respectively. The energy when the hydrogen is
at position $(-0.4\textrm{\AA},-0.4\textrm{\AA},-0.4\textrm{\AA})$
relative to the mid-point is set to a reference point of zero for
both the DFT and EVB calculations. \label{fig:EnergyPlanes1-1}}

\par\end{centering}

\end{figure}

\section{Results\label{sec:5Results}}

We have parametrised the EVB model against reference configurational
energies calculated from a DFT approach using a particle swarm optimisation
(PSO) fitting scheme, the details of which are described in the Appendix.
We focus in this section on investigating various features of the
parametrised model.

Figure \ref{fig:EnergyPlanes1-1} shows the potential energy of a
sulphuric acid/water system as a function of the position of a transferring
hydrogen atom, comparing the performance of the EVB model with a set
of reference data. It is important to recognise that the EVB scheme
does not operate for hydrogen positions beyond a distance $\pm0.3\textrm{\AA}$
in the $\mathbf{r}_{{\rm OO}}$ direction starting from $\frac{1}{2}(\mathbf{r}_{{\rm O}1}+\mathbf{r}_{{\rm O}2})$,
the mid-point between the oxygens atoms involved in the hydrogen bond,
and so energies at hydrogen positions labelled $\pm0.4\textrm{\AA}$
in the $\mathbf{r}_{{\rm OO}}$ direction are for unmixed bonding
and are not affected by the EVB parametrisation scheme. The comparison
demonstrates an overall correspondence between the EVB model and the
DFT results. The scheme is limited in that it is only active for hydrogen
positions lying in an ellipsoidal spatial region around the mid-point
between the oxygens, that it only represents two possible bonding
states per proton and that ultimately it is underpinned by a classical
potential fit to the reference data and an empirical form for the
off-diagonal term. Nevertheless, the EVB model seems to capture the
shape of the energy surface experienced by the hydrogen atom.

\noindent
\begin{figure}
\begin{centering}
\includegraphics[width=1\columnwidth]{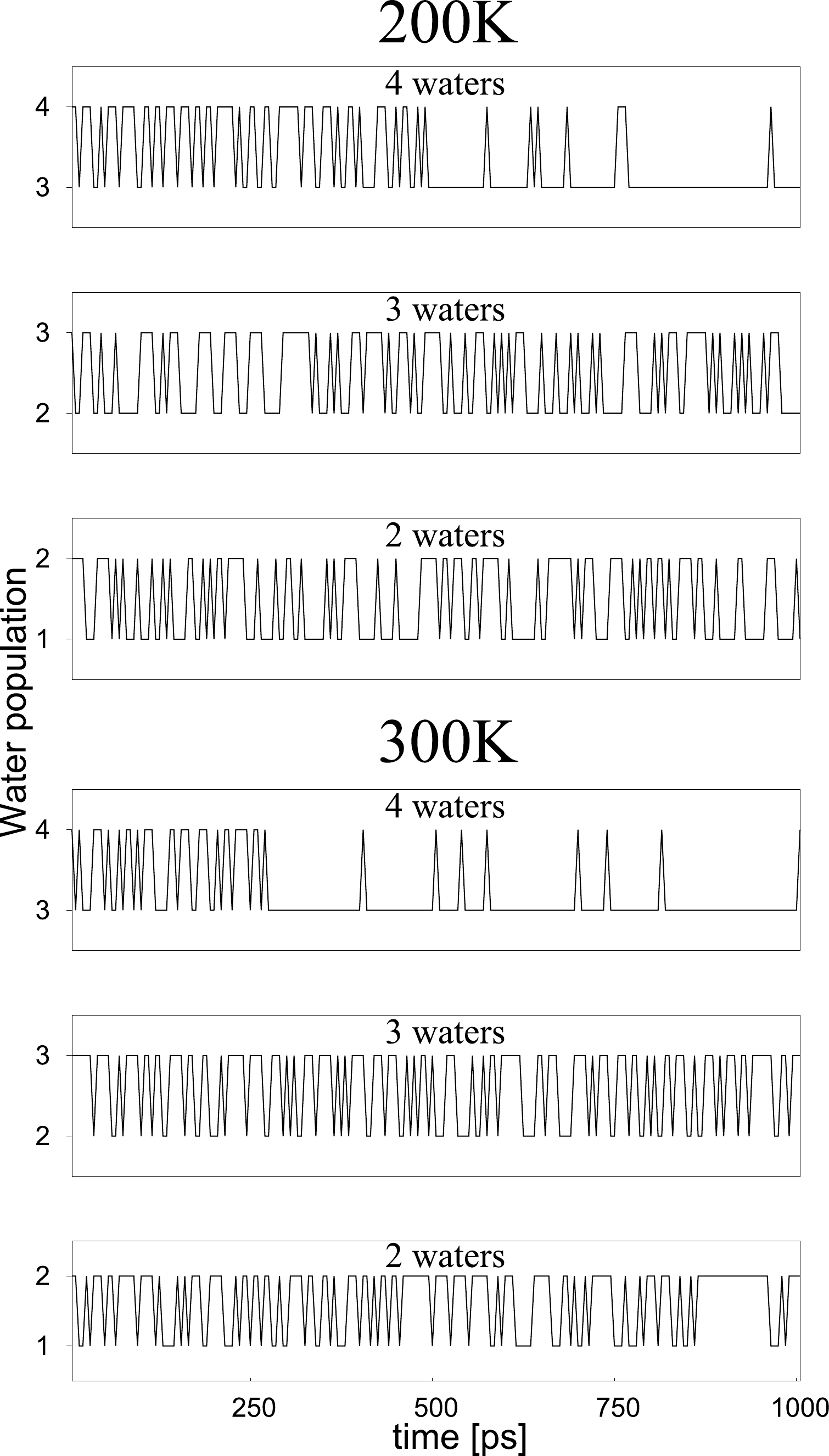}
\par\end{centering}

\protect\caption{\label{fig:Water_populations}Evolution in the population of water
molecules in simulations starting with the values indicated above
each frame. A decrease of one in the water population indicates an
ionisation event $\mathrm{[H_{2}SO_{4}]+[H_{2}O]_{n}\rightarrow[HSO_{4}]^{-}+[H_{3}O]^{+}+[H_{2}O]_{n-1}}$,
and an increase by one refers to the reverse of this reaction. The
simulations were performed at $\mathrm{200}$ K (top) and $\mathrm{300}$
K (bottom).}
\end{figure}

Next, we demonstrate how the propensity for proton transfer from an
acid molecule is affected by the extent of the surrounding water network.
Figure \ref{fig:Water_populations} tracks the population of water
molecules in various simulations of $\mathrm{[H_{2}SO_{4}]+[H_{2}O]_{n}}$
with $\mathrm{n=2-4}$, over a time period of $\mathrm{1}$ ns after
a $\mathrm{20}$ ps equilibration period. The simulations employed
a Langevin thermostat with target temperatures of $\mathrm{300\: K}$
or $\mathrm{200}$ K using a modified version of the DL\_POLY $\mathrm{4.03}$
program \citep{Todorov2006}. There was no constraint on the centre
of mass motion or the rotation of the clusters. The results show that
at $\mathrm{300}$ K there is very little propensity for the proton
to transfer to the closest water and remain there more than momentarily
in the $\mathrm{{\rm n}=2}$ and $\mathrm{{\rm n}=3}$ cases. The
hydrogen bond rarely fluctuates to such an extent that the ground
state configuration consists of a bisulphate/hydronium rather than
the sulphuric acid/water bonding pattern. However, for the $\mathrm{{\rm n}=4}$
case, the system spends the majority of its time in the dissociated
state. Proton transfers are more frequent and quasistable. When the
$\mathrm{{\rm n}=4}$ simulation is run at 200 K, however, we see
that it remains undissociated for a longer period of time suggesting
that the dissociated state is less stable at lower temperature, which
is to be expected. These conclusions are in agreement with work performed
at zero temperature using a higher level of theory \citep{Arrouvel2005,Re1999,Bandy1998},
according to which the first dissociation event of a sulphuric acid
molecule occurs when hydrated by between $\mathrm{3}$ and $6$ water
molecules.

When running the EVB model at 300 K we observe changes in configuration
that involve intermolecular bond making and breaking, characteristic
of the expected liquid-like nature of the system at relatively high
temperatures. We also observe the Grotthuss mechanism where protons
shuffle around the water network \citep{Paxton14}.

\section{Conclusions\label{sec:6Conclusion}}

A two-state EVB model for the sulphuric acid, bisulphate, hydronium
and water system has been presented, which is based upon the MS-EVB3
and SCI-MS-EVB models developed for hydronium/water alone \citep{Schmitt1998,Day2002}.
The approach essentially provides an interpolation between two energy
surfaces corresponding to the choices of bonding pattern available
to a proton in a particular spatial region between a bisulphate and
a water. It involves a Hamiltonian represented in a basis of bonding
patterns, with diagonal terms based on classical force fields, and
off-diagonal terms constructed empirically and fitted to higher level
calculations.

The limitation of the basis to two states implies that the model does not
match the state of the art in the description of protonated water systems \citep{Knight12} but
it provides the simplest scheme for representing proton transfer between sulphuric acid and water species,
the focus of our attention here and a new application of the EVB framework. An extension to include more
elaborate basis sets would be possible.

The parametrisation of the off-diagonal terms has been performed using
the particle swarm optimisation (PSO) technique, with reference to
DFT data for specific cases of sulphuric acid/water proton transfer
events. The resulting scheme has been integrated into a modified version
of the DL\_POLY 4.03 MD code \citep{Todorov2006}. A quantitative
comparison of the potential energy surface for the transferring proton
within a geometry-optimised static configuration of one sulphuric
acid and one water molecule gives good agreement. The level of hydration
required for the first ionisation or dissociation of a single sulphuric
acid molecule at $\mathrm{300\: K}$ has been investigated, revealing
that a system with four water molecules is best described by a complex
mixture between neutral and ionised bonding. At $\mathrm{200\: K}$
the four water case remains predominantly neutral, indicating that
an increase in temperature promotes proton transfer, as would be expected.
In highly hydrated structures, the proton becomes very mobile within
the water network.

The main motivation for developing this model is to provide a tool
for the fast simulation of cluster structures. The key performance
details are as follows. The run time for a 1 ns simulation with six
water molecules and one acid is $\mathrm{430\: seconds}$ on one Intel
Core i5-460M processor and it is, therefore, easily possible to generate
trajectories several tens of nanoseconds in length for quite substantial
systems using this model. The parametrisation procedure and tests
we have reported give us confidence that the physics of proton transfer
has been well captured by the scheme. Our computationally inexpensive
method is suitable for performing calculations of the thermodynamic
properties of clusters of water and sulphuric acid molecules at a
range of temperatures relevant to the atmosphere. We shall report
these results in future publications.

\section*{Acknowledgements}

SMK was supported in part by the U.S. Department of Energy, Office of Science, Office of Basic Energy Sciences, Division of Chemical Sciences, Geosciences, and Biosciences; JLS and IJF were supported by the IMPACT scheme at University College London (UCL). We acknowledge the UCL Legion High Performance Computing Facility, and associated support services together with the resources of the National Energy Research Scientific Computing Center (NERSC), which is supported by the U.S. Department of Energy under Contract No. DE-AC02- 05CH11231.  JLS thanks Dr. Gregory Schenter, Dr. Theo Kurt\'{e}n and Prof. Hanna Vehkam\"{a}ki for important guidance and discussions.



\appendix

\section{Parameter fitting procedure\label{sec:Fitting-procedure}}

The original MS-EVB3 model was parametrised to describe hydronium/water
proton transfers \citep{Wu2008}. We chose to employ slightly different
classical force fields for the water molecule and hydronium ion, requiring
us to revisit this parametrisation. The SPC/EF potential was employed
for water (see Loukonen \emph{et al.} \citep{Loukonen2010}) as well
as a modified version of the Morse potential in the MS-EVB3 model
(see \ref{sub:Underlying-classical-potential}). Three parameters
in the MS-EVB3 off-diagonal term for hydronium/water were refitted:
these were $R_{{\rm OO}}^{0}=2.7$\foreignlanguage{english}{\AA },
$P=0.4$ and $D_{{\rm OO}}=2.65$\foreignlanguage{english}{\AA{}}
and for this system we use $\tau_{x}=\tau_{yz}=0.3$.

Two further proton transfer reactions are possible in the system under
consideration, namely sulphuric acid/water and sulphuric acid/bisulphate
but we disregard the latter as it is likely to be rare. We parametrised
the new off-diagonal terms for sulphuric acid/water EVB mixing using
the particle swarm optimisation (PSO) scheme \citep{Banks2007b,Banks2007a}.
This technique has been employed in a variety of areas (reviewed in
reference \citep{Poli2008}) and we consider it to be sufficiently
powerful for use in potential fitting as well.

\begin{figure}
\begin{centering}
\includegraphics[width=1\columnwidth]{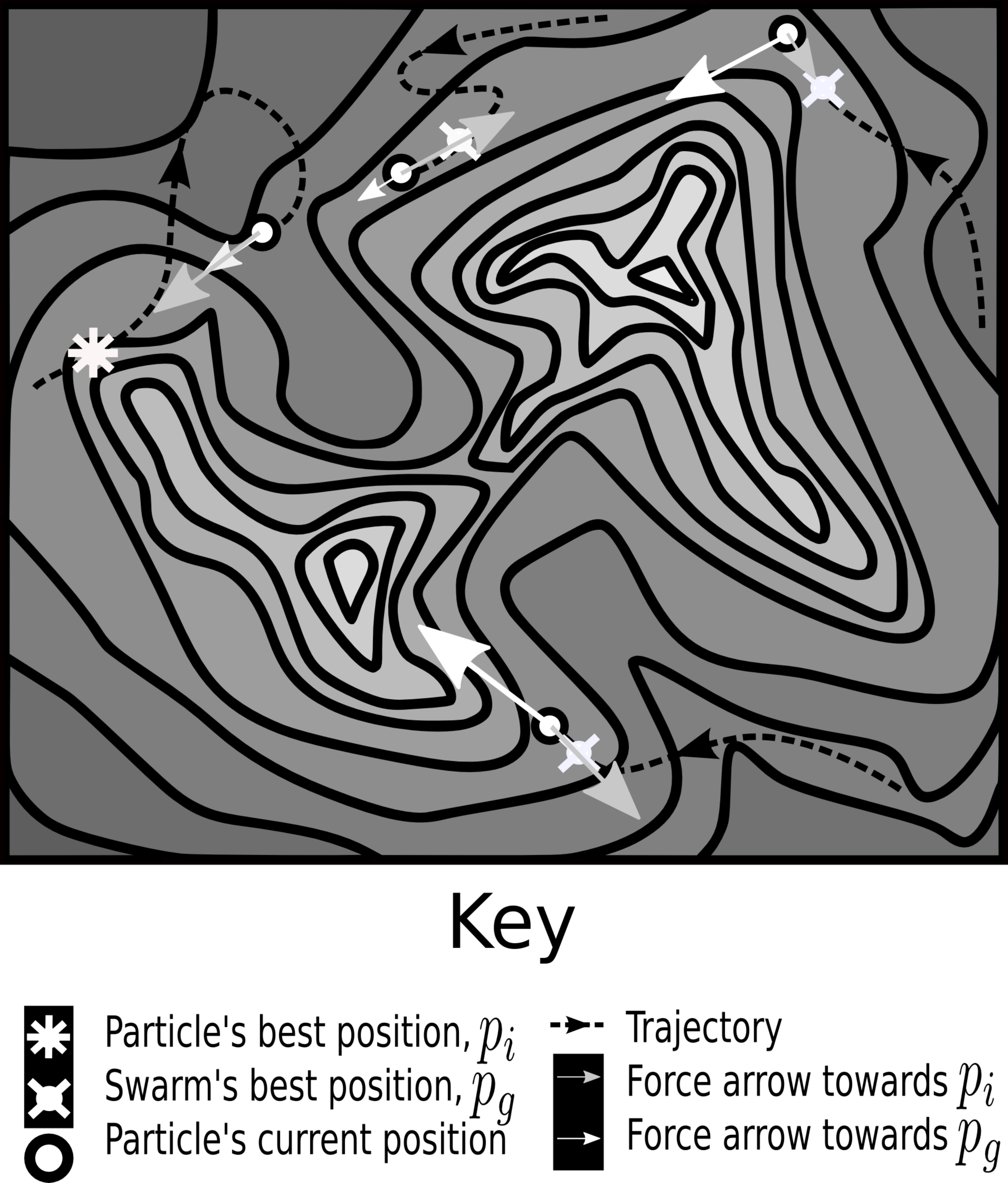}
\par\end{centering}

\protect\caption{Snapshot from a four `particle' PSO procedure to optimise two parameters
of a model. The darker shaded regions of parameter space represent
a poor fit between model and data, while lighter shades represent
a good fit (as determined by Eq. (\ref{eq:PSOFit})). The particles
follow trajectories through parameter space influenced by `forces'
directed towards individual ($p_{i})$ and collective ($p_{g})$ best
fit positions. The strength of the forces is influenced by noise introduced
into the swarm through variables $\beta_{1}$ and $\beta_{2}$.\label{fig:PSODia}}
\end{figure}

The PSO method is used to perform a force matching parametrisation
against DFT data. It employs a swarm of `particles', each representing
a choice of a set of parameters $\{\mathbf{\alpha}\}$. The goodness
of fit is assessed using a function, $f$, defined as
\begin{equation}
f(\{\mathbf{\alpha}\})=\sum\left(F_{{\rm ref}}-F_{{\rm model}}(\{\alpha\})\right)^{2},\label{eq:PSOFit}
\end{equation}
where $F_{{\rm ref}}$ denotes the reference forces for an atomic
configuration, and $F_{{\rm model}}(\{\alpha\})$ represents the model
forces for the same configuration, given parameters $\{\mathbf{\alpha}\}$.
The square deviations are summed over a series of configurations to
provide a global assessment of the fit between model and reference
data.

The particles are allowed to explore parameter space according to
a simple set of evolution equations for each parameter as described
by Shi and Eberhart \citep{Shi1998}:

\begin{equation}
\begin{array}{c}
v_{t+1}^{i}=\omega v_{t}^{i}+\varphi_{1}\beta_{1}(p_{i}-x_{t}^{i})+\varphi_{2}\beta_{2}(p_{g}-x_{t}^{i})\\
\\
x_{t+1}^{i}=x_{t}^{i}+v_{t+1}^{i}
\end{array}\label{eq:PSO-EqOfMotion}
\end{equation}

\noindent where $x_{t}^{i}$ and $v_{t}^{i}$ represent the `position'
and `velocity', respectively, of particle $i$ at iteration $t$.
The terms proportional to $\varphi_{1}$ and $\varphi_{2}$ are `forces'
designed to guide the particle towards the positions in parameter
space representing the previously found best fit parameter sets obtained
locally by the particle in question ($p_{i}$) and globally by the
entire swarm ($p_{g}$). $\beta_{1}$ and $\beta_{2}$ are uniformly
distributed random numbers in the range $0\leq\beta_{1,2}\le1$ designed
to introduce stochasticity into the search. The values of $\varphi_{1}$
and $\varphi_{2}$ represent the relative importance that is given
to the local and global best fits. The $\omega$ parameter is introduced
to allow a further flexibility in the algorithm, to be described later.
Figure \ref{fig:PSODia} illustrates the evolution of a four particle
swarm as it explores a two dimensional parameter space in search of
a minimum in the fitting function.

The PSO approach requires data to which the model is to be fitted
and also a numerical implementation specifying the number of particles
in the swarm and the number of independent searches performed, amongst
other matters.

\noindent
\begin{figure}
\begin{centering}
\includegraphics[width=1\columnwidth]{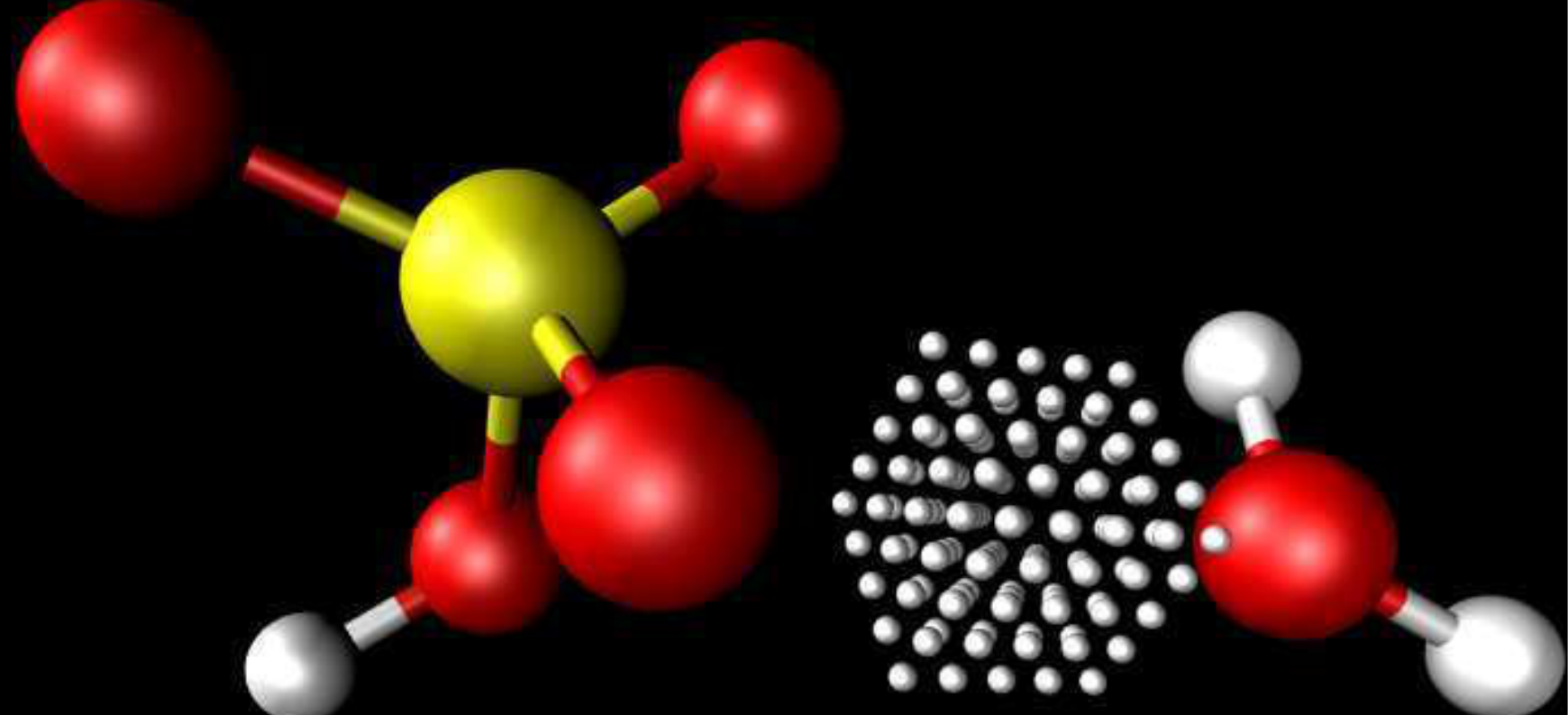}
\par\end{centering}

\protect\caption{\label{fig:555GridImage}The grid of hydrogen positions in configuration
I-n (in the terminology of reference \citep{Re1999}) used to generate
reference force data for the PSO fitting procedure.}
\end{figure}

\subsection{Reference data\label{sub:subsub:Fitting-data}}

The data against which the EVB model has been fitted was obtained
using density functional theory (DFT) based on the PBE \citep{Perdew1996}
functional with a plane wave basis set, a $\mathrm{550}$ eV cut off
and a ${\rm 15\:\hbox{\AA}}$ box. The CASTEP $\mathrm{5.5}$ code
\citep{Clark2005} was used to determine forces on atoms in a given
set of configurations. The configurations used in the dataset comprised:
\begin{enumerate}
\item Two molecular dynamics runs for trihydrated and tetrahydrated single
sulphuric acid molecules in configurations described as Config H and
SAQH in Stinson \emph{et al.} \citep{stinson14}.
\item A number of relaxed configurations designated I-n, II-n-a, III-n-a,
III-n-b and III-i-a according to Re \emph{et al.} \citep{Re1999}.
For each of these structures a proton was selected for transfer between
a sulphuric acid and a water molecule. This proton was positioned
on a grid of points in a cube centred at the mid-point between the
oxygens, $\frac{1}{2}(\mathbf{r}_{{\rm O}1}+\mathbf{r}_{{\rm O}2})$,
with one axis of the cube parallel to the $\mathbf{r}_{{\rm OO}}$
vector, and the other two axes orthogonal but arbitrary in direction.
Grid points were separated by $\mathrm{0.2\textrm{\AA}}$ giving a
total of 125 points per cube. Figure \ref{fig:555GridImage} illustrates
the grid of points for the I-n configuration. For each grid point
an energy and force calculation was performed.
\end{enumerate}
\noindent From this reference dataset, $\mathrm{400}$ configurations
were randomly selected for use in the PSO procedure.

\subsection{Implementation of the PSO method\label{subsubsub:Application-of-the-PSO-Method}}

The PSO method was run $\mathrm{200}$ times using a swarm of $\mathrm{50}$
particles which was chosen as a compromise between computational expense
and statistical noise. Each simulation was performed for $\mathrm{50}$
steps which was found to be sufficient for the parameters to have
converged. During the simulation, the value of $\omega$ in Eq. (\ref{eq:PSO-EqOfMotion})
was linearly scaled from $\mathrm{0.9}$ down to $\mathrm{0.4}$ as
suggested by Banks \emph{et al.} \foreignlanguage{english}{\citep{Banks2007a,Banks2007b}}.
$\varphi_{1}$ and $\varphi_{2}$ were both set equal to two.

There are ten parameters of the model to be optimised, and details
of the range searched for each parameter are given in Table \ref{tab:PSO-parameters}.
Three parameters were not treated in this way. $r_{{\rm OO}}^{0}$
was set to zero as it was noted to be redundant. The parameter $\gamma$
was set to the value used in the MS-EVB3 model (see Table \ref{tab:MSEVB3Values}).
The parameter $\tau_{x}$ was chosen to be $0.3$. The EVB model parameters
were taken to be the averages of the parameter sets $\{p_{g}\}$ obtained
from the $\mathrm{200}$ runs; this was seen as a fair compromise
between computational expense and accuracy. The resulting optimised
values are given in Table \ref{tab:PSO-parameters}.

\begin{table}
\caption{\label{tab:PSO-parameters}Table of EVB parameters which were fitted using the PSO method. The `Range' \textit{\emph{considered}} for each parameter is shown. During each PSO simulation the maximum permitted velocity of parameter change was linearly scaled from max to min values  in each case. Units are as given in Table \ref{tab:MSEVB3Values}, with $\tau_{yz}$ dimensionless and $\Delta$ in kcal/mol.}
{\begin{tabular}{cccccc}
\toprule
Param. & Range (max) & Range (min) & Vel. (max) & Vel. (min) & Fit\\
\colrule
$V_{\rm const}^{ij}$ & 0.0 & -100.0 & 10.0 & 0.2 &  -72.20998		    \\
$P$ & 1.0 & 0.01 & 0.1 & 0.002 &  0.50743							 \\
${\rm k}$ & 30.0 & 1.0 & 3.0 & 0.06 & 15.64862							  \\
$D_{\rm OO}$ & 8.0 & 0.1 & 0.8 & 0.016 &  4.18888					     \\
$\beta$ & 8.0 & 0.1 & 0.8 & 0.016 &  2.37963						  \\
$R_{\rm OO}^{0}$ & 8.0 & 0.1 & 0.8 & 0.008 &  2.46345					 \\
$P'$ & 30.0 & 1.0 & 3.0 & 0.06 &  15.41834							\\
$\alpha$ & 30.0 & 5.0 & 3.0 & 0.06 &  17.42592						\\
$\tau_{yz}$ & 15.0 & 0.5 & 1.5 & 0.03 & 11.55606						 \\
$\Delta$ & 800.0 & 400.0 & 20.0 & 0.05 &  558.40454				   \\
\botrule
\end{tabular}}
\end{table}

\noindent
\end{document}